# Loss compensation symmetry in dimers made of gain and lossy nanoparticles

**VASILY KLIMOV**[1,2,3,*] **ILYA ZABKOV**[1], **DMITRY GUZATOV**[5], **AND ALEXEY VINOGRADOV**[1,4,6]

[1]*All-Russia Research Institute of Automatics - 127055, Moscow, Russia*
[2]*P.N. Lebedev Physical Institute, Russian Academy of Sciences - 119991, Moscow, Russia*
[3]*National Research Nuclear University «MEPhI» - 115409, Moscow, Russia*
[4]*Moscow Institute of Physics and Technology (State University) - 141700, Dolgoprudny, Russia*
[5]*Yanka Kupala State University of Grodno - 230023, Grodno, Belarus*
[6]*Institute for Theoretical and Applied Electromagnetics RAS - 125412, Moscow, Russia*
klimov256@gmail.com

**Abstract:** The eigenoscillations in a two-dimensional dimer made of gain and lossy nanoparticles were investigated within exact analytical approach. It was shown that there are eigenoscillations for which all Joule losses are exactly compensated by the gain. Among such solutions, there are solutions with a new type of symmetry, which we refer to as Loss Compensation Symmetry (LCS) as well as well-known PT (parity-time) symmetric solutions. The modes with LCS unlike PT symmetric ones allow one to achieve full loss compensation with significantly less gain that in the case of PT symmetry. This effect paves way to the new loss compensation methods in optics.

## References and links

## 1. Introduction

In many cases for deep understanding of physics as well as for many hi-tech applications, one needs to consider systems with small inner losses (high-Q systems). Eigenoscillations in such systems are most pronounced, and one can think about new possible applications of scientific results. Systems with small inner losses do not improve the performance of corresponding devices only quantitatively but often show a qualitatively new behaviour. In particular, when the light is scattered by a single nanoparticle with small inner losses, an electric quadrupole mode can dominate in scattering spectrum over the dipole one [1]. Moreover, unusual modes such as left-handed plasmonic [2] or chiral modes [3,4] can be observed in the system with small losses. In dimers, i.e. systems consisting of two particles, in the case of small losses and small distances between the particles, an essentially new type of localized oscillations ("plasmonic molecules") can be seen [5-7].

High-Q electromagnetic resonances can be easily discovered in the IR frequency range, where almost transparent materials with high permittivity (silicon, germanium, etc.) exist. In the MIR frequency range, the low losses regime can be achieved with phonon-polariton resonances (SiC and other [8]) or with graphene [9,10]. At visible frequencies band, finding of high-Q oscillations in nanoparticles is a more complicated task because typical for this range plasmon resonances have quite big optical losses. Hence, many interesting theoretical predictions can be hardly observed.

The idea of loss compensation with making use of gain materials is widely discussed nowadays. Substantial progress is achieved in loss compensation in metamaterials [11-13]. Among all systems with loss compensation, parity-time (PT) symmetric systems exhibit very interesting effects, such as symmetry breaking and power oscillations [14-20]. Let us remind that in optics, PT symmetry means that permittivity of system satisfies the following relation: $\varepsilon(x,y,z) = \varepsilon^*(-x,y,z)$ [14].

In this work, we will present results of investigation of loss compensation in a general case of a dimer made of particles of the same shape but with arbitrary relation between gain and loss. For clearness, we will consider two-dimensional (2D) oscillations in a dimer made of two 2D nanoparticles with permittivities $\varepsilon_G = \varepsilon' - i\varepsilon_G''\,(\varepsilon_G'' > 0)$, $\varepsilon_L = \varepsilon' + i\varepsilon_L''\,(\varepsilon_L'' > 0)$ for gain and lossy nanoparticle, respectively. Geometry of the problem is shown at Fig. 1.

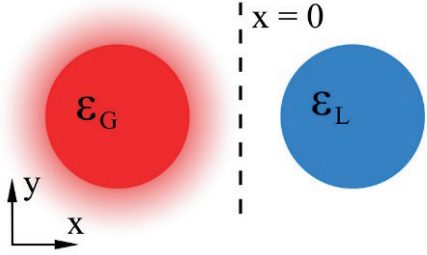

Fig. 1. Geometry of the problem under consideration. A dimer consisting of two 2D nanoparticles with equal radii. The imaginary part of permittivity of the right particle is positive $\mathrm{Im}\,\varepsilon_L > 0$ (lossy material), whereas for the left particle $\mathrm{Im}\,\varepsilon_G < 0$ (gain material). The real parts are equal $\mathrm{Re}\,\varepsilon_L = \mathrm{Re}\,\varepsilon_G$.

Let us note that we consider general reasoning, and specific realization of the dimer is out of scope of this work. However, several possible approaches to obtain gain nanoparticles are known [17, 21].

## 2. Material independent eigenvalues and eigenmodes

By an exact analytical calculation of eigenmodes, we have shown that full loss compensation can be obtained not only for PT symmetric systems, i.e. when $\varepsilon_G'' = \varepsilon_L''$, but also for the nonsymmetrical case $\varepsilon_G'' \neq \varepsilon_L''$. We will refer to the eigenmodes in such asymmetrical systems as to modes with Loss Compensation Symmetry (LCS). Such modes allow one to achieve a full compensation of optical losses even in the case $\varepsilon_G'' \ll \varepsilon_L''$.

We define eigenoscillations of such a system as nontrivial solutions of Maxwell's equations without any sources. As spectral parameters, we consider permittivity values of the system ($\varepsilon'$, $\varepsilon_G''$, $\varepsilon_L''$). Such an approach is very powerful because eigenvalues and eigenfunctions found within it do not depend on specific materials. The solution of specific problem with specific materials can be easily discovered on the base of known eigenfunctions and eigenvalues. In this case, Eigen frequencies can be found from specific dispersion relation between permittivity and frequency.

It is important to note that making use of classical Maxwell's equations is correct only when the distance between nanoparticles in dimer is bigger than few nanometers. Otherwise, quantum effects are to be taken into consideration [22-24].

For small in comparison with wavelength particles (nanoparticles), retardation effects are small and a quasistatic approach can be used. In this case, to find eigenoscillations one should solve the Laplace equation for potential satisfying the conditions of continuity for potential and normal component of electric displacement vector on all boundaries:

$$\mathbf{E} = -\nabla\varphi,\ \Delta\varphi = 0, \qquad (1)$$

$$[\varphi]_S = 0,\ [\varepsilon\,\partial\varphi/\partial n]_S = 0, \qquad (2)$$

where **E** is the electric field vector, $\varphi$ is the potential, $\nabla$ is the spatial gradient, $\Delta$ is the Laplacian, and S is the nanoparticles surfaces. Let us note that by considering the quasistatic problem, the radiation losses are not taken into account. It is very important, that $\varepsilon$ as eigenvalues of (1) and (2) are negative numbers, which can be proved by potential theory [25].

In the case of 2D oscillations (which do not depend on the $z$ coordinate), to solve (1), (2) it is convenient to introduce bipolar coordinates $\theta, \eta$, which are related to Cartesian coordinates $x, y$ by the following relations [26]:

$$\eta = \operatorname{atanh}\left(\frac{2fx}{f^2 + x^2 + y^2}\right), \theta = \operatorname{atan}\left(\frac{2fy}{f^2 - x^2 - y^2}\right), \tag{3}$$

In bipolar coordinate system, an arbitrary solution of the Laplace equation (1) can be expressed as a sum of the functions $e^{\pm n\eta} e^{\pm in\theta}, n = 1, 2...$ [26]. In the case of equal size particles, their boundaries are defined by the equations $\eta = \pm \eta_0 (\eta_0 > 0)$, which correspond to the radius $f / \sinh \eta_0$ and the distance between the particles centres $2f \coth \eta_0$. The particles permittivities are $\varepsilon_G$ and $\varepsilon_L$ respectively, while the permittivity of the surrounding medium is taken to be $\varepsilon_m = 1$ (vacuum).

Potential distribution outside and inside the particles can be written in the form:

$$\varphi_G = ae^{n\eta}, \varphi_L = de^{-n\eta}, \varphi_{\text{out}} = be^{n\eta} + ce^{-n\eta}, \tag{4}$$

where $\varphi_G, \varphi_L, \varphi_{\text{out}}$ are the potentials inside gain and lossy particles and outside them, respectively. The unknown coefficients $a, b, c, d$ can be found by applying the boundary conditions (2). In (4) and below for briefness, we omit the factors $\cos(n\theta)$ or $\sin(n\theta)$ in potentials. Substituting (4) into (2), the following exact dispersion relation can be obtained:

$$(\varepsilon_L - 1)(\varepsilon_G - 1) = (\varepsilon_L + 1)(\varepsilon_G + 1) e^{4n\eta_0}, \tag{5}$$

which is valid for each azimuthal number $n$. If (5) is valid, the following important relations between coefficients in (4) can be easily obtained:

$$a/d = (1 - \varepsilon_L)(1 + \varepsilon_G)^{-1} e^{-2n\eta_0},$$
$$b/d = 0.5(1 - \varepsilon_L) e^{-2n\eta_0}, c/a = 0.5(1 - \varepsilon_G) e^{-2n\eta_0}. \tag{6}$$

It is also very important that one can strictly prove that any eigensolution of (1) satisfies the condition of full loss compensation, that is:

$$\int_{V_G} \varepsilon_G'' |\mathbf{E}(\mathbf{r})|^2 d\mathbf{r} = \int_{V_L} \varepsilon_L'' |\mathbf{E}(\mathbf{r})|^2 d\mathbf{r}, \tag{7}$$

where $V_G, V_L$ are the volumes of gain and lossy nanoparticles. Thereby if one attributes effective permittivity [27] to the whole system, its imaginary part will be equal to zero, so our system is similar to pseudo-Hermitian systems [28].

Let us note that in the case with retardation, an additional term appears in the right side of (7), which corresponds to radiative losses in the system. To satisfy (7), the value of $\varepsilon_G''$ will become bigger with the increasing of radiative losses. In this work, we make use of the quasistatic approximation where radiative losses are neglected.

First of all, let us consider the PT symmetric system with:

$$\varepsilon_G = \varepsilon' - i\varepsilon'', \quad \varepsilon_L = \varepsilon_G^*, \tag{8}$$

where $\varepsilon', \varepsilon''$ are real numbers. Substituting (8) into (5), one can obtain the following relation between the real and imaginary parts of permittivity:

$$\left[\varepsilon' + \coth(2n\eta_0)\right]^2 + \varepsilon''^2 = 1/\sinh^2(2n\eta_0). \tag{9}$$

As it follows from Eq. (9), the dispersion curve for PT symmetric system in coordinate space $\varepsilon', \varepsilon''$ is a circle with the radius $1/\sinh(2n\eta_0)$ and with the center at $\left[\varepsilon' = -\coth(2n\eta_0), \varepsilon'' = 0\right]$. If $\varepsilon'' < 1/\sinh(2n\eta_0)$, the PT symmetric system has PT symmetric eigenoscillations:

$$\varphi(x) = \varphi^*(-x). \tag{10}$$

The specific form of (10) is given below [see the structure of potential given by (18)]. However, if $\varepsilon'' > 1/\sinh(2n\eta_0)$, the PT symmetric system (8) has no eigenmodes at all. In literature, it is usually referred to as PT symmetry breaking [14].

For asymmetrical system ($\varepsilon_L \neq \varepsilon_G^*$), eigenmodes exist only for one specific value of the real part of permittivity (for fixed $n$), which is defined by geometrical parameters only:

$$\varepsilon'_{LCS} = -\left(e^{4n\eta_0} + 1\right)/\left(e^{4n\eta_0} - 1\right), \tag{11}$$

while the imaginary parts of the permittivity for gain and lossy nanoparticles can vary relative to each other, but their product is fixed:

$$\varepsilon''_{G,LCS}\varepsilon''_{L,LCS} = 4e^{4n\eta_0}/\left(e^{4n\eta_0} - 1\right)^2. \tag{12}$$

It follows from (12) that in our dimer, the full loss compensation is possible for an arbitrary value of the imaginary part of permittivity in a lossy particle. Moreover, the greater $\varepsilon''_{L,LCS}$ is, the smaller $\varepsilon''_{G,LCS}$ is required for full loss compensation, because they depend on each other inversely. This effect results from the asymmetry of the field distribution between gain and lossy particles, so when $\varepsilon''_{L,LCS}$ gets bigger, the amplitude of field in lossy particle decreases relative to the active particle (see Fig. 4).

One can show that the eigenmodes of an asymmetrical system with permittivities given by (11) and (12) have a nontrivial symmetry:

$$\varphi(x) = \varphi^*(-x)\left[\theta(-x)\sqrt{\varepsilon''_L/\varepsilon''_G} + \theta(x)\sqrt{\varepsilon''_G/\varepsilon''_L}\right], \quad |\eta| > \eta_0, \tag{13}$$

where $\theta(x)$ is Heaviside step function. We will refer to this symmetry as Loss Compensation Symmetry (LCS). For PT symmetric case ($\varepsilon''_L = \varepsilon''_G$), this symmetry is reduced to PT symmetry.

The full spectrum of eigenmodes of the system under consideration is shown in Fig. 2 for particles with the radius 10 nm and the distance between their centres 28 nm.

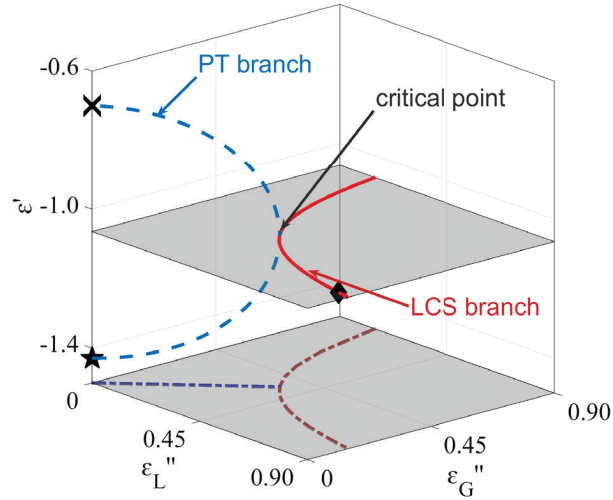

Fig. 2. Dispersion curves for 2D eigenoscillations in two particles with permittivities defined by $\varepsilon_L = \varepsilon' + i\varepsilon_L''$, $\varepsilon_G = \varepsilon' - i\varepsilon_G''$, corresponding to eigenmodes with $n=1$. The dashed blue line corresponds to PT symmetric modes, $\varepsilon_G'' = \varepsilon_L''$, while the solid red one corresponds to LCS modes. The projection of the solid and dashed lines on the bottom plane is shown by the dash-dotted lines.

As it can be seen in this figure, the spectrum of PT symmetric modes indeed lies in the plane $\varepsilon_G'' = \varepsilon_L''$, while the spectrum of LCS modes lies in the perpendicular plane $\varepsilon' = \text{const}$. It's worth of mentioning that the branches of the dispersion curve touch each other at one point (marked as the critical point in Fig. 2), allowing one to tell about spontaneous breaking of PT symmetry at this point and transition to solutions with LCS. The values of permittivity $\varepsilon_L^c$, $\varepsilon_G^c$ of particles for this critical point are defined only by geometrical parameters of the system and can be easily deduced from (11), (12):

$$\varepsilon_L^c = -\left(e^{4n\eta_0} + 1\right)/\left(e^{4n\eta_0} - 1\right) + i2e^{2n\eta_0}/\left(e^{4n\eta_0} - 1\right),$$
$$\varepsilon_G^c = \left(\varepsilon_L^c\right)^*. \qquad (14)$$

In Fig. 2, the eigenmodes are shown for $n=1$ only. Eigenmodes with higher azimuthal number $n$ will have the similar shape, and the real part of the permittivity of particles $\varepsilon'$ will tend to minus unity in accordance with (11).

Now let us consider spatial distribution of potentials for PT symmetric eigenmodes. PT symmetry of potential means that:

$$\varphi(x) = \varphi^*(-x). \qquad (15)$$

From (15) it follows immediately that the real part of the potential should be symmetric, while the imaginary part – antisymmetric. From (15), it also follows that in (4) one should put:

$$a = d^*. \qquad (16)$$

It can be demonstrated that (16) is true only when (8) is satisfied, i.e. the potential can have PT symmetry only if the system permittivity does have it itself. Extracting the real and imaginary parts of a, $a = a' + ia''$, from (6) and (16), one can obtain:

$$a'' = -i\frac{(F-1)}{(F+1)}a', \quad F = \frac{1-\varepsilon_L}{1+\varepsilon_G}e^{-2n\eta_0}, \tag{17}$$

where despite of an imaginary unit in (17), $a''$ remains real for all PT eigenmodes except $\varepsilon_G'' = \varepsilon_L'' = 0$. The potential at any point of PT symmetric branch can be written in the form:

$$\varphi = a'\varphi^s + ia''\varphi^a, \tag{18}$$

where the symmetric $\varphi^s$ and antisymmetric $\varphi^a$ parts of the potential are real and universal and can be found from (4) by the substitution:

$$a = 1, \, b = 0.5(1-\varepsilon_L)e^{-2n\eta_0}, \, c = 0.5(1-\varepsilon_G)e^{-2n\eta_0}, \, d = 1, \tag{19}$$

and

$$a = 1, \, b = -0.5(1-\varepsilon_L)e^{-2n\eta_0}, \, c = 0.5(1-\varepsilon_G)e^{-2n\eta_0}, \, d = -1, \tag{20}$$

respectively. For pure real permittivities (marked as cross and star on Fig. 2) either $a''$ or $a'$ are equal to zero, and eigenfunction becomes symmetric or antisymmetric relative to the parity transformation $x \to -x$. Spatial distribution of $\varphi^s$ and $\varphi^a$ is shown in Fig. 3.

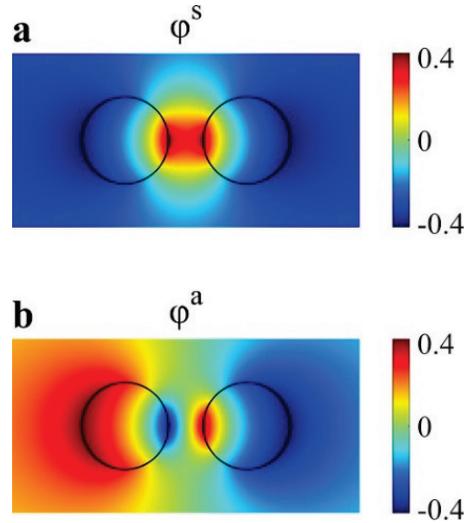

Fig. 3. Distribution of symmetric $\varphi^s$ and antisymmetric $\varphi^a$ part of potentials corresponding to PT eigenmodes (18) with $n=1$. The angle dependence is $\cos\theta$.

Now let us consider asymmetrical systems with $\varepsilon_L \neq \varepsilon_G^*$ and related LCS modes. As full loss compensation exists for any solution of (1), one can find an additional relation between $a$ and $d$ in (4):

$$\varepsilon_G''|a|^2 - \varepsilon_L''|d|^2 = 0. \tag{21}$$

By choosing of an appropriate arbitrary phase, the solution of (21) and (6) can be written as:

$$a = a' + ia'', \quad d^* = a/\kappa, \tag{22}$$

where

$$a'' = -i\frac{(F/\kappa - 1)}{(F/\kappa + 1)} a', \quad \kappa = \sqrt{\varepsilon_L'' / \varepsilon_G''}. \tag{23}$$

Here again, despite of an imaginary unit in (23), $a''$ remains real for all LCS eigenmodes. Other coefficients in (4) are defined by (6). For a general case (22), the LCS has a place for a potential inside the particles:

$$\varphi(x) = \varphi^*(-x)\left[\kappa\theta(-x) + \theta(x)/\kappa\right], |\eta| > \eta_0. \tag{24}$$

The potential outside the particles $|\eta| < \eta_0$ does not satisfy (24). Let us stress that equations (22), (23), (24) transform to (16), (17), (15) in the case of PT symmetric system because for that case $\kappa = \sqrt{\varepsilon_L'' / \varepsilon_G''} = 1$.

Spatial distribution of the potential for LCS mode is shown in Fig. 4 in a case of substantial difference between $|\varepsilon_G''|$ and $|\varepsilon_L''|$ in nanoparticles: $\varepsilon_G = -1.06 - 0.156i$ and $\varepsilon_L = -1.06 + 0.85i$.

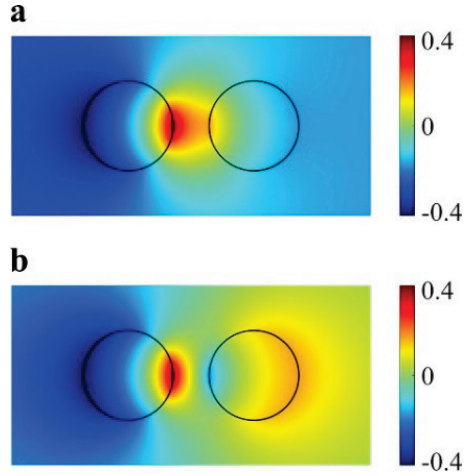

Fig. 4. The real (a) and imaginary (b) part of LCS eigenmode potential with $n = 1$. $\varepsilon_G = -1.06 - 0.156i$, $\varepsilon_L = -1.06 + 0.85i$ The point with corresponding parameters is shown as a rhomb at Fig. 2. The angle dependence is $\cos\theta$.

It can be seen from Fig. 4 that amplitude of the potential (and therefore field) in the left (gain) particle is much higher than in the right (lossy) one. Meanwhile, the modulus of the permittivity imaginary part in right particle is much higher than in the left one $\varepsilon_L'' = 0.85 >> \varepsilon_G'' = 0.156$. In Fig. 4, it is difficult to trace LCS (19). However, if one plots the distribution of Joule losses $\sim \varepsilon''(\mathbf{r})|\mathbf{E}(\mathbf{r})|^2$ in space (see Fig. 5), the LCS and full loss compensation becomes evident.

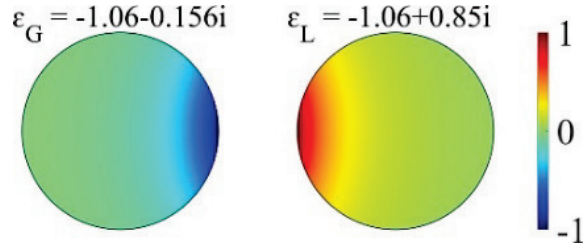

Fig. 5. Normalized value of Joule's losses in a dimer. The parameters are the same as in Fig. 4.

Let us note that we consider only case with equal real parts of permittivity of particles. If one would consider more general case with $\varepsilon_G = \varepsilon'_G - i\varepsilon''_G\,(\varepsilon''_G > 0)$, $\varepsilon_L = \varepsilon'_L + i\varepsilon''_L\,(\varepsilon''_L > 0)$ then 2D surface in 4D space with $\varepsilon'_G, \varepsilon''_G, \varepsilon'_L, \varepsilon''_L$ coordinates would be a solution.

## 3. Possibility of application to real systems

From Fig. 2 one can see that for "interesting" regime of LCS modes ($\varepsilon''_L \gg \varepsilon''_G$), a specific relation between the imaginary and real parts of the permittivity of a lossy cylinder is needed. Particularly for this regime, the imaginary part of the permittivity should be of the same order of magnitude as the real one with the absolute value of the real part close to 1. This condition can be satisfied, for example, for gold. However, a promising plasmonic material TiN (see e.g. [8, 29-31]) would do better because of a bigger permittivity imaginary part in the visible part of the spectrum. In Fig. 6, one can see both material independent PT and LCS modes for different radii and permittivities of real metals (TiN [32] and Au [33]).

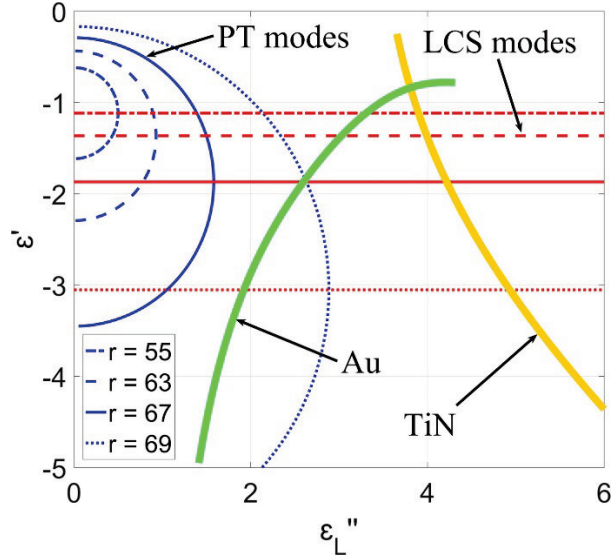

Fig. 6. The superposition of material independent PT (circles) and LCS (horizontal lines) modes for different radii (55, 63, 67, 69 nm) and permittivities of TiN (wavelengths 470 to 600 nm) and Au (wavelengths 470 to 540 nm). The distance between the cylinders is 140 nm.

In the case when the retardation effects should be taken into account, the electric field can be expressed through the vector $\mathbf{A}$ and the scalar $\varphi$ potentials as:

$$\mathbf{E} = ik_0\mathbf{A} - \nabla\varphi, \quad \mathrm{div}\mathbf{A}=0. \tag{25}$$

where k0 is the wavenumber for vacuum. In this case instead of (1), (2), one can use the following formulation of the eigenmode problem:

$$\Delta \varphi = 0,$$
$$[\varphi]_S = 0, \, [\varepsilon \partial \varphi / \partial n]_S = ik_0 [\varepsilon \mathbf{A}_n]_S. \tag{26}$$

From the wave equation for the vector potential:

$$\Delta \mathbf{A} + k_0^2 \varepsilon \mathbf{A} = -ik_0 \varepsilon \nabla \varphi, \tag{27}$$

one can see that $\mathbf{A} = O(k_0)$. Therefore, (26) can be simplified to:

$$\Delta \varphi = 0,$$
$$[\varphi]_S = 0, \, [\varepsilon \partial \varphi / \partial n]_S = O(k_0^2). \tag{28}$$

From (28), it follows that the retardation effects for nanoparticles result only in small changes in the eigenvalue problem (1,2), and therefore the solutions we have found for PT and LCS modes are valid in the case with retardation, too. The numerical simulations of full system of Maxwell equations confirm this fact. Therefore, new LCS modes can be observed at telecom and visible wavelengths.

## 4. Conclusion

In conclusion, the problem of 2D eigenoscillations in a dimer made of gain and lossy nanoparticles has been solved analytically. Found eigenoscillations appear when the whole optical losses in the first particle are compensated by amplification of the optical fields in the second particle. It was demonstrated that together with PT symmetric modes in PT symmetric system ($\varepsilon_G'' = \varepsilon_L''$), there are new types of modes for the asymmetrical system with $\varepsilon_G'' \neq \varepsilon_L''$. We refer to them as modes with Loss Compensation Symmetry (LCS). These modes require a small absolute value of the permittivity imaginary part in an active particle to compensate the optical losses in a lossy particle with a high value of the permittivity imaginary part. This feature of LCS modes paves the way to new methods of loss compensation. In addition, all modes are characterized by mixing of symmetric and antisymmetric modes which exist in nanoparticles without losses.

### Funding

The research has been supported by the Advanced Research Foundation (Contract No. 7/004/2013-2018). Authors acknowledge financial support from the Russian Foundation for Basic Research (Grants No. 14-02-00290 and No. 15-52-52006).